 \documentclass[epjST]{svjour}
\usepackage{amsmath,amsfonts,amssymb,graphicx,color,times,bbm,psfrag,dsfont}
\usepackage[T1]{fontenc}
\usepackage{times}
\usepackage{url}

\begin{document}

\title{
Different models of gravitating Dirac fermions in optical lattices
}

\author{Alessio Celi \thanks{\email{alessio.celi@gmail.com}}}

\institute{ICFO-Institut de Ci\`{e}ncies Fot\`{o}niques, The
Barcelona
  Institute of Science and Technology, 08860 Castelldefels
  (Barcelona), Spain
}

\abstract{
In this paper I construct the naive lattice Dirac Hamiltonian describing the propagation of fermions in a generic 2D optical metric 
for different lattice and flux-lattice geometries. First, I apply a top-down constructive approach that we first proposed in [Boada {\it et al.,New J. Phys.} {\bf 13} 035002 (2011)]  
to the honeycomb and to the brickwall lattices. I carefully discuss how gauge transformations that generalize momentum (and Dirac cone) shifts in the Brillouin zone in the Minkowski 
homogeneous case can be used in order to change the phases of the hopping. In particular, I show that lattice Dirac Hamiltonian for Rindler spacetime in the honeycomb and brickwall 
lattices can be realized by considering real and isotropic (but properly position dependent) tunneling terms. For completeness, I also discuss a suitable formulation of Rindler Dirac Hamiltonian 
in semi-synthetic brickwall and $\pi$-flux square lattices (where one of the dimension is implemented by using internal spin states of atoms as we originally proposed in [Boada {\it et al.,Phys. Rev. Lett. } {\bf 108} 133001 (2012)]
and [Celi {\it et al.,Phys. Rev. Lett. } {\bf 112} 043001 (2012)]).
}
\maketitle

\section{Introduction}

In the last decade the emergence of Dirac fermions in condensed matter and low energy physics has become 
central in Physics due to graphene revolution \cite{graphene_nobel} and due to the discovery of topological insulators \cite{topology_nobel,Kane05}.  
Indeed, many of the amazing properties of graphene, namely, being a high-mobility semiconductor with zero cyclotron mass at half filling \cite{CastroNeto09},
can be derived by simple tightbinding analysis \cite{Wallace47} and explained in terms of the existence of Dirac cones that determine the relativistic nature of quasi-particle
excitations at low energy.
On the other hand, topological properties and emergence of edge states can be also explained in terms of Dirac operators \cite{Schnyder08,Hasan10}. 
The latter explains also the existence of Dirac semimetals in 3D materials which has been recently demonstrated in \cite{Xu15} (for a very recent review see \cite{Chiu16}).  
Building on the lesson of graphene, the emergence of relativistic particles can be forced by generating Dirac cones in the energy bands of properly chosen lattice systems,
as, for instance, ultracold atoms in bichromatic \cite{Salger11}, hexagonal \cite{Soltan-Panahi11,Duca15} 
and brickwall lattices \cite{Tarruell12} but also in artificial lattice Dirac systems such as
nano-patterned 2D electron gases, photonic crystals, micro-wave
lattices \cite{Polini13} or polaritons \cite{Jacqmin14}. Note that Dirac cones can be generated also in continuous systems like trapped 
ultracold gases by artificial laser induced spin-orbit coupling \cite{Juzeliunas07,Unanyan10,Lin11,Galitski13,Huang16}. 
While the existence of Dirac cones is completely kinematic and it is property of single particle solutions, and, thus, 
completely unrelated to particle statistics, only in fermionic systems Dirac cones at the proper filling control the low-energy dynamics as in graphene,
dynamics that can be probed for instance by Landau-Zener transitions \cite{Tarruell12,Uehlinger13}.  

The range of interesting phenomena that can be observed in graphene or simulated in artificial Dirac systems (in any dimensions) is enormous \cite{Lewenstein12,Celi16}.
As observed for instance in \cite{Mazza11}, by changing the properties under discrete symmetries of the lattice model that hosts Dirac points it is in principle possible 
to achieve topological insulators in all the classification classes. For instance, the celebrated Haldane model \cite{Haldane88} recently experimentally demonstrated with ultracold atoms in
an optically shaken brickwall lattice \cite{Jotzu14} can be interpreted as a realization of lattice Dirac Hamiltonian without doubling due to the breaking of the chiral symmetry \cite{Nielsen81}.    
Furthermore, systems governed by the Dirac Hamiltonian display also anomalous Hall conductivity \cite{Novoselov05,Zhang05,Geim07,Goldman09,Watanabe10} 
and puzzling properties like Klein tunneling \cite{Klein29} and zitterbewegung \cite{Schrodinger30,David10},
phenomena that are accessible preferably or uniquely with graphene  \cite{Cserti06,Rusin09,Katsnelson06} (or graphene like compounds, see \cite{Zawadzki11})
or artificially engineered systems as in ultracold neutral atoms \cite{Otterbach09,Vaishnav08,Merkl08,Zhang10,Lepori10,LeBlanc13}, trapped ions \cite{Casanova10,Gerritsma11,Lamata07,Gerritsma10,Casanova11}, 
photons \cite{Longhi10,Longhi10b,Dreisow10}, conductor quantum wells \cite{Schliemann05}, and circuit QED \cite{Pedernales13,Liu14}.

More generally, quantum simulators of Dirac Hamiltonians allow for the simulation of high energy physics phenomena like neutrino oscillations \cite{Lan11,Noh12,Wang15}, axion 
electrodynamics \cite{Bermudez10} or Schwinger effect \cite{Szap11}, Dirac fermions in interactions \cite{Cirac11},
and in principle relativistic Dirac fermions are required in phenomenological oriented quantum simulation of quantum field theory \cite{Casanova11b,Semiao12,Jordan12}, in particular of 
lattice gauge theories, subject that has received recently considerable attention,
due to the fascinating perspective of understanding phase diagram and dynamics of Abelian \cite{Buchler05,Zohar11,Zohar12,Tagliacozzo12,Banerjee12} 
and non-Abelian \cite{Tagliacozzo13,Banerjee13,Zohar13} gauge theories with ultracold atoms \cite{Stannigel14,Kasper15,Dutta16} and other table-top experiments \cite{Hauke13,Martinez16,Yang16,Marcos13} 
(for reviews see \cite{Wiese13,Zohar15}). 
Note that in parallel also classical simulation of gauge theory based on tensor networks have received great attention 
\cite{Tagliacozzo11,Banuls13,Liu13,Buyens14,Tagliacozzo14,Silvi14,Kuhn14,Haegeman15,Banuls15,Kuhn15,zohar2015fermionic,Pichler16,zohar2016building,Dittrich16,zohar2016projected,Silvi16}.

Last but not least, emerging Dirac fermions offer the possibility of observing the exotic and intriguing phenomena  due to the interplay between gravity and field 
theory \cite{Birrell_Davies}. The simulation of the Hawking radiation \cite{Hawking75} and of the Unruh effect \cite{Unruh76} does not certainly require relativistic fermions \cite{Barcelo05} (see also \cite{Volovik03,Gibbons14}).
Indeed, it can be performed, for instance, with relativistic bosonic quasiparticle like phonons in a BEC \cite{Garay00,Fedichev.03,Fedichev.04,Fedichev.04b,Retzker.PRL.08,Westbrook.12,Steinhauer.14,Westbrook.15,Marino.16}
--for  a very recent experiment and discussions about it, see \cite{Steinhaurer16} and \cite{discussions}, respectively-- or in a ion trap 
\cite{Alsing05,Schutzhold07}, with photons \cite{Philbin.08,Belgiorno.PRL.10,Unruh.PRL.11,Unruh.PRD.12,Finazzi.14}  or just with classical analogue as 
waves in water \cite{Unruh.81,Weinfurtner.PRL.11,Weinfurtner.13}. Quantum simulators of Dirac fermions in curved spacetimes as we first proposed in \cite{Boada2010b} and later considered also in \cite{Minar15,Koke16}
allow in principle not only to study single particle phenomena in different dimensions as we have done recently for the Unruh effect \cite{Rodriguez16} but also to systematically include interactions
in addition to tuning the spacetime geometry.

In fact, the propagation of Dirac fermions in curved spacetime was first considered in graphene by Cortijo and Vozmediano \cite{Cortijo07a,Cortijo07b} 
for quantifying the effect of ripples on the conduction and the density of carriers of  graphene sheet rather than as a tool for quantum simulation. 
Although the extrinsic metric in graphene corresponds to spatial deformations of the Minkowski metric, Iorio and Lambiase \cite{Iorio12} noted that 
by very specifically shaping the graphene sheet and exploiting the Weyl invariance of conductivity \cite{Iorio11} it would be possible to observe
Hawking-Unruh effect in such sample. Indeed, the effective metric for the graphene carriers becomes conformally equivalent to the one 
of a black hole, while their Whightman correlation function is invariant under this conformal transformation and, thus, display the same thermal behavior
as in presence of the black hole. This approach based on conformal transformation has some difficulties pointed in \cite{Cvetic12} by Cvetic and Gibbons who
argued that there is a fundamental geometric obstacle to obtaining a model that extends all the way to the black hole horizon with a finite graphene sheet 
(for a more advance discussion on the properties of optical metrics and of their relation with cosmological and holographic solutions can be found in \cite{Cvetic16}).
  Then, Iorio and Lambiase replayed by showing that a way out to the problem above exists, and different conformal maps can be considered, which allow to reach the horizon
  on a finite lattice at the price of a non-thermal correction in the Wightman response function \cite{Iorio14}. Note that also different embedding of 
  the graphene can be considered, in particular it has been shown very recently by Cariglia {\it et al.} that deformed bilayer graphene admits a natural embedding in 
  4D curved spacetime and that conductivity is controlled by the curvature \cite{Cariglia16}.   

It is worth to notice that quantum simulation of curved spacetime in optical lattices cannot follow the same route as in graphene, essentially because the laser
beams of the former cannot be bended, and another strategy has to be consider.
There are indeed two different ways of simulating the motion in artificial curved background. 
The first, which can be called geometrical, is to consider the $D-$dimensional system, for instance $D=2$, as a hyper-surface in $D+1$ flat space. 
If the embedding is not trivial the  (extrinsic) induced  metric is not. This is the case for graphene-based materials or graphene itseft. 
The electronic properties in presence of defects of ripples may be described in the long wavelength approximation as Dirac fields propagating in such spacetime metrics. 
The second, which we developed in \cite{Boada2010b} and can be regarded as Newtonian, 
is to incorporate the effect of gravity in the dynamics by changing the Hamiltonian governing the system. 
Roughly speaking, the metric is treated similarly to a background gauged field. 
In \cite{Boada2010b}, we showed that for a special class of metric the corresponding Dirac Hamiltonian on a square lattice 
can be obtained by modulating the intensity of the hopping in each site of the lattice, accordingly to the metric.

An advantage of our approach  is that is top-down, 
in the sense that the natural procedure is to derive that lattice Hamiltonian of interest 
starting from the continuous Hamiltonian and discretizing it in position space. 
In this paper, I show the power of this method. In Sect. \ref{sect:II} I derive the graphene-like lattice Hamiltonian for a honeycomb lattice 
in presence of a background metric in the class studied in \cite{Boada2010b}, 
and an Abelian gauge field, Sect. \ref{sect:III}. 
Apart few subtleties related to the non-orthogonality of the lattice generating vectors, 
the derivation goes on similar lines as for a square lattice, with the difference that the tunneling terms come out 
generically complex. In fact, I show in Sect. \ref{sect:IV} that, with properly chosen gauge transformations that generalize the momentum shifts
of the Brillouin zone in Minkowski space, it is possible to achieve real and isotropic tunneling terms in
paradigmatic example of Rindler spacetime. Then, in Sect. \ref{sect:V} I repeat the same construction for deformed hexagonal lattice, that is the brickwall
lattice. In particular, I show that Dirac Hamiltonian in Rindler spacetime can be obtained again by simply shaping the intensity of the tunneling term to have linear slope.
Furthermore, I provide the implementation of the brickwall as a semi-synthetic lattice, that is with one real dimension and one synthetic (extra-)dimension obtained by coupling 
the spin states of fermionic atoms, as we originally proposed in \cite{Boada12} and applied to the simulation of integer quantum Hall effect and of the corresponding chiral edge states in
\cite{Celi14} (for the experimental realization of the proposal see \cite{Stuhl15,Mancini15,Celi15}, for other applications of synthetic lattices see 
\cite{Grass14,Boada15,Grass15,Mugel16,Price15,Zeng15,Zhang15,Ozawa16,Wall16,Bilitewski16,Yuan16,Meier16,Livi16,Suszalski16,Meier16b,Ghosh16,An16,Barbarino16,Price16,Ghosh16b,Ghosh16c,Ozawa16b,Anisimovas16,Saito16}). 
In Sect. \ref{sect:VI} I give the implementation of the 
Dirac Hamiltonian in curved spacetimes on a bipartite square lattice, which it is also known (in its Minkowski version) as $\pi$-flux Hamiltonian \cite{Kogut.75,Affleck.88,Lim.09} 
because, in order to restore the braiding property
of a Dirac spinor around a plaquette, an artificial magnetic flux of $\pi$ is required. Finally, I conclude with some final remarks in Sect. \ref{sect:VII}.      

Before starting a disclaimer: the Hamiltonian coming out of our procedure is the naive Hamiltonian, 
as it is affected by the doubling of the poles. However, this is not even a disease here as it does not spoil, for instance, the properties of Unruh effect and related phenomena.

\section{The straightforward Dirac Hamiltonian on the hexagonal lattice is not the graphene one}\label{sect:II}

Let me start by the continuous Hamiltonian to be discretized. Following \cite{Boada2010b}, for a metric background of the form 
\begin{equation}
ds^2= - J({\bf r})^2 dt^2 + dx^2+dy^2\,,\label{metric}
\end{equation}
the corresponding Dirac Hamiltonian can be written simply as
\begin{equation}
H=\frac i2 \int {\rm dx} {\rm dy} J({\bf r}) \left(\partial_p \psi^\dagger({\bf r}) \sigma_p \psi({\bf r}) - \psi^\dagger({\bf r}) \sigma_p \partial_p \psi({\bf r}) \right)\,.\label{contham}  
\end{equation}       
To fix the notation, $ {\bf r}=(x,y)$, $p=x,y$, $\psi({\bf r})=\left(\begin{array}{c}a({\bf r})\\b({\bf r})\end{array}\right)$ is a spinor and the $\sigma_p$ are the usual Pauli matrices
\begin{equation}
\sigma_x=\left(\begin{array}{cc} 0&1\\1&0\end{array}\right),\ \ \ \ \ \ \ \ \sigma_y=\left(\begin{array}{cc} 0&-i\\i&0\end{array}\right).
\end{equation}

In this notation the Hamiltonian \eqref{contham} can be rewritten as
\begin{equation}
H=\frac i2 \int {\rm dx} {\rm dy} J({\bf r}) \left((\partial_x - i \partial_y)  a^\dagger({\bf r}) b({\bf r}) - a^\dagger({\bf r}) (\partial_x - i \partial_y) b({\bf r}) \right)+ {\rm H.c.}\,.\label{contham1}  
\end{equation}


\begin{figure}[hb!]\sidecaption
\resizebox{0.35\columnwidth}{!}{%
\includegraphics{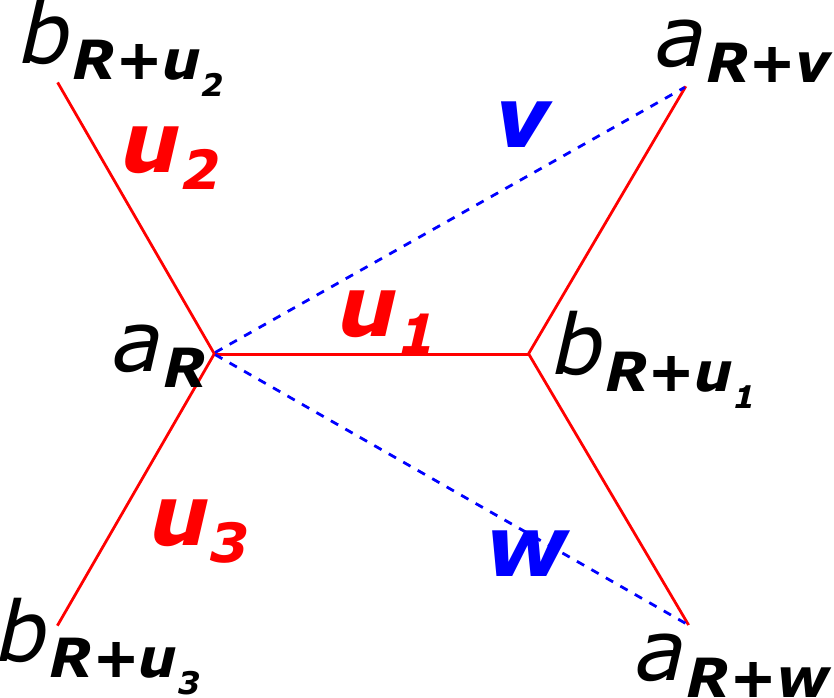} }
\caption{The honeycomb lattice, indicated in red, as bipartite lattice formed by two triangular sublattices generated by the vectors ${\bf v}$ and ${\bf w}$, 
indicated in blue and given in the main text.
The displacement between the two sublattice is given by the links of the honeycomb lattice, ${\bf u_j}$, $j=1,2,3$, for instance of ${\bf u_1}$.
The links add up to zero, $\sum_j {\bf u_j}=0$, and are chosen here to be of unit length. In terms of the generators they read,
${\bf u_1} =\frac{{\bf v} + {\bf w}}3$, ${\bf u_2} =\frac{{\bf v} - 2{\bf w}}3$, and ${\bf u_3} =\frac{{\bf w} - 2{\bf v}}3$.  
}
\label{fig:1}       
\end{figure}

The second step is to determine the shape of the lattice we are interested in, see Fig. \ref{fig:1}. The honeycomb lattice where the links are 
\begin{equation}
{\bf u_1}=(1,0),\ \ \ \ \ {\bf u_2}=(-\frac 12,\frac{\sqrt 3}2),\ \ \ \ \ {\bf u_3}=(-\frac 12,-\frac{\sqrt 3}2),
\end{equation}
is the superposition of two Bravis lattices generated by ${\bf v}={\bf u_1}-{\bf u_3}=(3/2,\sqrt 3 /2)$ and ${\bf w}={\bf u_1}-{\bf u_2}=(3/2,-\sqrt 3 /2)$ and displaced by ${\bf u_1}$ (or any other link vectors). 

The third step is to substitute the derivatives of the spinor in $x$ and $y$ with finite differences of the spinor components' on the lattice points,
 and to substitute the integral with a sum over the lattice points.

There are two issues. 
The first is that there are many equivalent ways of decomposing a displacement parallel to $x$ and $y$ 
in terms of the vectors ${\bf v}$ and ${\bf w}$ (${\bf u_1}$, ${\bf u_2}$,  and ${\bf u_3}$), 
i.e.,  many collections of points can be chosen to compute the same derivative. 
The second is that the spinor components $a({\bf r})$ and $b({\bf r})$ do not live on the same site.
 This second problem is related to the first one as in first approximation for instance
\begin{equation}
b( {\bf R})\simeq b( {\bf R}+{\bf u_1}) -\partial_x b( {\bf R}+{\bf u_1}),
\end{equation}
where ${\bf R}=  m {\bf v} + n {\bf w}$, $m,n\in {\mathbb Z}$ is a generic point in the sublattice occupied by the fermion $a$.

One possible way of proceeding is to consider the following relations that are valid at first order
\begin{align}
\partial_x a^\dagger({\bf R}) &\simeq\frac 13 \left( a^\dagger_{{\bf R} + {\bf v}} + a^\dagger_{{\bf R} + {\bf w}} - 2 a^\dagger_{\bf R} \right) \cr
\partial_y a^\dagger({\bf R}) &\simeq\frac 1{\sqrt 3} \left( a^\dagger_{{\bf R} + {\bf v}} - a^\dagger_{{\bf R} + {\bf w}}  \right) \cr
\partial_x b({\bf R}) &\simeq\frac 13 \left( 2 b_{{\bf R} + {\bf u_1}} - b_{{\bf R} + {\bf u_2}} - b_{{\bf R} + {\bf u_3}} \right) \cr
\partial_y b({\bf R}) &\simeq\frac 1{\sqrt 3} \left( b_{{\bf R} + {\bf u_2}} - b_{{\bf R} + {\bf u_3}} \right) \cr
b({\bf R}) &\simeq\frac 13 \left( b_{{\bf R} + {\bf u_1}} + b_{{\bf R} + {\bf u_2}} + b_{{\bf R} + {\bf u_3}} \right) \cr
           &\simeq\frac 13 \left(-2 b_{{\bf R}+{\bf v}+{\bf u_1}} + b_{{\bf R}+{\bf v}+{\bf u_2}}+ 4 b_{{\bf R}+{\bf v}+{\bf u_3}}\right) \cr
           &\simeq\frac 13 \left(-2 b_{{\bf R}+{\bf w}+{\bf u_1}} + 4 b_{{\bf R}+{\bf w}+{\bf u_2}}+ b_{{\bf R}+{\bf w}+{\bf u_3}}\right)\,. \label{linexp}
\end{align}
The expressions above can be derived by noticing that at first order, by indicating with $c$ generically $a$, $b$, $a^\dagger$, $b^\dagger$ and with
${\bf S}$ a generic vector, $c({\bf R} + {\bf S}) \simeq  c({\bf R}) + {\bf S} \cdot {\bf \nabla}\, c({\bf R})$, which by linearity implies 
\begin{equation}
\sum_l d_l \,c({\bf R} + {\bf S}_l) = \left(\sum_l d_l\right)\, c({\bf R}) + \left(\sum_l d_l {\bf S}_l \right) \cdot {\bf \nabla}\,  c({\bf R}) .
\end{equation}
Thus, in order to obtain the expressions for the derivatives along $\hat x$ ($\hat y$), one has simply to require (or check in this case) that  $\sum_l d_l=0$
and $\sum_l d_l {\bf S}_l = \hat x$ ($\hat y$). Instead, the expressions for $b({\bf R})$ are obtained by requiring that $\sum_l d_l=1$ and  $\sum_l d_l {\bf S}_l = 0$.
As explained before the set of displacements $\{ {\bf S}_l \}$ is chosen in order to construct the desired tightbinding model.

By using the relations \eqref{linexp} we can discretize the Hamiltonian \eqref{contham}.
In particular, we notice that tricky binomials like $a( {\bf R}+{\bf v})^\dagger b( {\bf R})$ and 
$a( {\bf R}+{\bf w})^\dagger b( {\bf R})$ can be expressed in terms of nearest-neighbor tunnelings  
\begin{align}
a( {\bf R}+{\bf v})^\dagger b( {\bf R})&\rightarrow \frac 13 a_{\bf R+{\bf v}}^\dagger \left(-2 b_{{\bf R}+{\bf v}+{\bf u_1}} + b_{{\bf R}+{\bf v}+{\bf u_2}}+ 4 b_{{\bf R}+{\bf v}+{\bf u_3}}\right)\cr
a( {\bf R}+{\bf w})^\dagger b( {\bf R})&\rightarrow \frac 13 a_{\bf R+{\bf w}}^\dagger \left(-2 b_{{\bf R}+{\bf w}+{\bf u_1}} + 4 b_{{\bf R}+{\bf w}+{\bf u_2}}+ b_{{\bf R}+{\bf w}+{\bf u_3}}\right). \label{binrule}
\end{align}
By exploiting that the sum over ${\bf R}$ is running over the whole plane (which is a good approximation for a sufficiently large lattice) we can replace for instance 
$\sum_{\bf R} a_{{\bf R} +{\bf v}}^\dagger b_{{\bf R} +{\bf v}+{\bf u_1}}$ with  $\sum_{\bf R} a_{{\bf R }}^\dagger b_{{\bf R} +{\bf u_1}}$ and get as lattice Hamiltonian
\begin{equation}
H=-i \sum_{\bf R} \sum_{j=1}^3  J_{{\bf u_j},{\bf R}} a_{{\bf R }}^\dagger b_{{\bf R} +{\bf u_j}} + {\rm H.c.}\,,\label{latham}
\end{equation} 
where 
\begin{align}
J_{{\bf u_1},{\bf R}}&= \frac 19 \left[\left(J_{{\bf R}- {\bf v}} +J_{{\bf R}- {\bf w}} + 4 J_{{\bf R}}\right) +\sqrt 3 i \left(J_{{\bf R}- {\bf v}}-J_{{\bf R}- {\bf w}}\right)\right],\cr
J_{{\bf u_2},{\bf R}}&= - \frac 1{18}\left[\left(J_{{\bf R}- {\bf v}} +4 J_{{\bf R}- {\bf w}} + J_{{\bf R}}\right) -\sqrt 3 i \left(J_{{\bf R}- {\bf v}}- 4 J_{{\bf R}- {\bf w}}- 3 J_{{\bf R}}\right)\right],\cr
J_{{\bf u_3},{\bf R}}&= - \frac 1{18}\left[\left(4 J_{{\bf R}- {\bf v}} +J_{{\bf R}- {\bf w}} + J_{{\bf R}}\right) -\sqrt 3 i \left( 4 J_{{\bf R}- {\bf v}} - J_{{\bf R}- {\bf w}}+ 3 J_{{\bf R}}\right)\right].
\end{align}
Notice that the global phase $-i$ can be eliminated, for instance, by a phase redefinition of the $a$'s, $i\, a_{\bf R} \to a_{\bf R}$, which obviously implies  $-i\, a^\dagger_{\bf R} \to a^\dagger_{\bf R}$.  
As we constructed our lattice Hamiltonian to be graphene like, it is worth to consider the propagation in the Minkowski metric, i.e.,  for a spatially constant hopping, 
$J_{\bf R}=J$. The hopping over the different links reduce to
\begin{equation}
J_{\bf u_1}\rightarrow \frac 23 J,\ \ \ \ J_{\bf u_2}\rightarrow \frac 23 e^{-i \frac{2\pi}3} J,\ \ \ \ J_{\bf u_3}\rightarrow \frac 23 e^{i \frac{2\pi}3} J\label{constantj}.
\end{equation}
  At the first sight, the outcome is quite surprising as the hopping is not the same for the different links as in graphene and not even {\it real}. 
  Furthermore, it is not possible to remove the phases by a global phase transformation \footnote{It should be specified that the phases can not be removed by a global phase 
  transformation in the $(x,y)$  coordinate systems. Indeed, as discussed below the hopping phases correspond to a pure gauge configuration, i.e.,  to a gauge field with zero flux. 
  This means that it exists a gauge transformation that removes the gauge field. 
  This also implies that this transformation is just a global phase transformation in momentum space.} 
  of the spinor $\psi$, or equivalently by a redefinition of the Pauli matrices by a rotation around the $z$-axis. 
  However, there is nothing wrong with the lattice Hamiltonian we have found. 
  Indeed, as a check, we can verify the existence of two Dirac points, 
  which are equivalent to the graphene-like model but have a different location in the Brillouin zone.  
  For the hopping (\ref{constantj}), the condition that Hamiltonian in momentum space is zero, 
  $\sum_{j=1}^3 J_{\bf u_j} e^{i {\bf k}\cdot{\bf u_j}} =0$, implies that as inequivalent Dirac points can be chosen the origin, 
  ${\bf k}=(0,0)$ ($\sum_{j=1}^3 J_{\bf u_j}=0$), and   ${\bf k}=\frac{4 \pi}{3\sqrt 3}(\frac 12 ,-\frac{\sqrt 3}2)$, which lays on the frontier of the Brillouin zone.
  Thus, this configuration corresponds to a displacement in momentum of the Brillouin zone of ${\bf K}_D=\frac{4 \pi}{3\sqrt 3} (0,1)$, which is equivalent
  to the following local gauge transformation in momentum space
  \begin{align}
  a_{\bf R} \to e^{-i {\bf K}_D \cdot {\bf R}} a_{\bf R},\cr
  a^\dagger_{\bf R} \to e^{i {\bf K}_D \cdot {\bf R}} a^\dagger_{\bf R},\cr
  b_{\bf R} \to e^{-i {\bf K}_D \cdot {\bf R}} b_{\bf R},\cr
  b^\dagger_{\bf R} \to e^{i {\bf K}_D \cdot {\bf R}} b^\dagger_{\bf R},  
  \end{align}
  The transform above implies for the tunnelings
  \begin{equation}
  J_{{\bf u_j},{\bf R}} \to e^{-i {\bf K}_D \cdot {\bf u_j}} J_{{\bf u_j},{\bf R}},
  \end{equation}
  which gives the phases in \eqref{constantj} as ${\bf K}_D \cdot {\bf u_1}=0$, and ${\bf K}_D \cdot {\bf u_2}=- {\bf K}_D \cdot {\bf u_3} = \frac{2\pi}3$. 
  Note that the module of the tunneling we have obtained for the Minkowski case, $\frac 23 J$, it is nothing more than the relation between the tunneling and Fermi velocity in
  graphene that for the lattice spacing we have chosen is precisely equal to $J$.
  Coming back to a generic curved spacetime described by the metric \eqref{metric}, what we find suggests that our lattice model (\ref{latham}) is {\it gauge equivalent} to the gravitational deformation of graphene-like model. 
  In the next section we will show how this relation can be made explicit by the inclusion of the gauge field coupling in the continuous Hamiltonian we start with.      
  
 \section{Gauge\&Gravity coupled Dirac Hamiltonian on a hexagonal lattice}\label{sect:III}

 I am going to repeat the same exercise as in the previous section for Dirac charged particles coupled to a  gauge field and moving in the metric (\ref{metric}). 
 Note that this exercise has some relation with the debate \cite{deJuan12} on whether the effect of ripples and other in graphene is better accounted by gravitational distortion or 
 by the presence of gauge fields. For a comprehensive discussion we refer the reader to \cite{Arias15} where a unique relation between the space curvature and the magnetic field induced by ripples in graphene
 is established.
 
 By the gauge choice $A_0=0$ \footnote{This gauge choice it is always possible, but in presence of a non zero electric field implies a time dependent vector potential. 
 In what follows we restrict to a purely magnetic configuration, $A_p=A_p({\bf r})$.}, 
 this is equivalent to consider the Hamiltonian
\begin{equation}
H=\frac i2 \int {\rm dx} {\rm dy} J({\bf r}) \left((\partial_p-i A_p({\bf r})) \psi^\dagger({\bf r}) \sigma_p \psi({\bf r}) - \psi^\dagger({\bf r}) \sigma_p (\partial_p + i A_p({\bf r})) \psi({\bf r}) \right)\,,\label{gaugecontham}  
\end{equation}   
which in components reads
\begin{multline}
H=\frac i2 \int {\rm dx} {\rm dy} J({\bf r}) \left[(\partial_x - i \partial_y -i (A_x{\bf r}) -i A_y{\bf r}) )  a^\dagger({\bf r}) b({\bf r})\right. \cr 
                                                  \left. - a^\dagger({\bf r}) (\partial_x - i \partial_y + i (A_x{\bf r}) -i A_y{\bf r})) b({\bf r}) \right]+ {\rm H.c.}\,.\label{gaugecontham1}  
\end{multline}
%
%
The only new ingredient that we have to add to the recipe is
\begin{align}
\left(\partial_x - i A_x \right) a^\dagger({\bf R}) &\simeq\frac 13 \left( e^{-i \int_{\bf R} A_v} a^\dagger_{{\bf R} + {\bf v}} + e^{-i \int_{\bf R} A_w} a^\dagger_{{\bf R} + {\bf w}} - 2 a^\dagger_{\bf R} \right) \cr
\left(\partial_y - i A_y \right) a^\dagger({\bf R}) &\simeq\frac 1{\sqrt 3} \left( e^{-i \int_{\bf R} A_v} a^\dagger_{{\bf R} + {\bf v}} - e^{-i \int_{\bf R} A_w} a^\dagger_{{\bf R} + {\bf w}}  \right) \cr
\left(\partial_x + i A_x \right) b({\bf R}) &\simeq\frac 13 \left( 2 e^{i \int_{\bf R} A_{u_1}} b_{{\bf R} + {\bf u_1}} - e^{i \int_{\bf R} A_{u_2}} b_{{\bf R} + {\bf u_2}} 
                                                                   - e^{i \int_{\bf R} A_{u_3}} b_{{\bf R} + {\bf u_3}} \right) \cr
\left(\partial_y + i A_y \right) b({\bf R}) &\simeq\frac 1{\sqrt 3} \left( e^{i \int_{\bf R} A_{u_2}} b_{{\bf R} + {\bf u_2}} - e^{i \int_{\bf R} A_{u_3}} b_{{\bf R} + {\bf u_3}} \right) \, , \label{gaugelinexp}
\end{align}
where the expression $\int_{\bf R} A_{S}$ for a generic vector ${\bf S}$ is a short cut for the line integral $\int_0^1{\rm d}l\,  {\bf S} \cdot {\bf A}({\bf R}+ l\,{\bf S} )$.
Again the relations above can be checked by Taylor expanding the right hand sides at first order and by exploiting the linearity of the scalar products, ${\bf S} \cdot {\bf A}({\bf R})$.


By using the relation \eqref{linexp} and \eqref{gaugelinexp} we get again a Hamiltonian of the form (\ref{latham}) with the hopping of the form
\begin{align}
J_{{\bf u_1},{\bf R}}&= \frac 19\left[\left(J_{{\bf R}- {\bf v}} e^{-i \int_{{\bf R}-{\bf v}} A_v} 
                                            + J_{{\bf R}- {\bf w}} e^{-i \int_{{\bf R}-{\bf w}} A_w} + J_{{\bf R}} \left(1 + 3 e^{-i \int_{{\bf R}} A_{u_1}}\right) \right)\right.\cr 
                     &\ \ \ \ \  \left.  +\sqrt 3 i \left(J_{{\bf R}- {\bf v}} e^{-i \int_{{\bf R}-{\bf v}} A_v} - J_{{\bf R}- {\bf w}} e^{-i \int_{{\bf R}-{\bf w}} A_w} \right)\right],\cr
J_{{\bf u_2},{\bf R}}&= - \frac 1{18}\left[\left(J_{{\bf R}- {\bf v}} e^{-i \int_{{\bf R}-{\bf v}} A_v} 
                                            + 4 J_{{\bf R}- {\bf w}} e^{-i \int_{{\bf R}-{\bf w}} A_w} + J_{{\bf R}} \left(-2 + 3 e^{-i \int_{{\bf R}} A_{u_2}}\right) \right)\right.\cr 
                     &\ \ \ \ \  \left.  -\sqrt 3 i \left(J_{{\bf R}- {\bf v}} e^{-i \int_{{\bf R}-{\bf v}} A_v} - 4 J_{{\bf R}- {\bf w}} e^{-i \int_{{\bf R}-{\bf w}} A_w} - 3 J_{{\bf R}} e^{-i \int_{{\bf R}} A_{u_2}} \right)\right],\cr
J_{{\bf u_3},{\bf R}}&= - \frac 1{18}\left[\left(4 J_{{\bf R}- {\bf v}} e^{-i \int_{{\bf R}-{\bf v}} A_v} 
                                            + J_{{\bf R}- {\bf w}} e^{-i \int_{{\bf R}-{\bf w}} A_w} + J_{{\bf R}} \left(-2 + 3 e^{-i \int_{{\bf R}} A_{u_3}}\right) \right)\right.\cr 
                     &\ \ \ \ \  \left.  -\sqrt 3 i \left( 4 J_{{\bf R}- {\bf v}} e^{-i \int_{{\bf R}-{\bf v}} A_v} - J_{{\bf R} - {\bf w}} e^{-i \int_{{\bf R}-{\bf w}} A_w} + 3 J_{{\bf R}} e^{-i \int_{{\bf R}} A_{u_3}} \right)\right].
                     \label{gaugetun}
\end{align}

%

As a final exercise we look for the pure gauge configuration that reproduces the graphene like Hamiltonian for $J_{\bf R}=J$. Under this condition the above expressions reduce to
\begin{align}
J_{{\bf u_1},{\bf R}}&= \frac 19\left[\left( e^{-i \int_{{\bf R}-{\bf v}} A_v} 
                                            +  e^{-i \int_{{\bf R}-{\bf w}} A_w} + \left(1 + 3 e^{-i \int_{{\bf R}} A_{u_1}}\right) \right)\right.\cr 
                     &\ \ \ \ \  \left.  +\sqrt 3 i \left( e^{-i \int_{{\bf R}-{\bf v}} A_v} -  e^{-i \int_{{\bf R}-{\bf w}} A_w} \right)\right] J,\cr
J_{{\bf u_2},{\bf R}}&= - \frac 1{18}\left[\left( e^{-i \int_{{\bf R}-{\bf v}} A_v} 
                                            + 4  e^{-i \int_{{\bf R}-{\bf w}} A_w} +  \left(-2 + 3 e^{-i \int_{{\bf R}} A_{u_2}}\right) \right)\right.\cr 
                     &\ \ \ \ \  \left.  -\sqrt 3 i \left( e^{-i \int_{{\bf R}-{\bf v}} A_v} - 4  e^{-i \int_{{\bf R}-{\bf w}} A_w} - 3  e^{-i \int_{{\bf R}} A_{u_2}} \right)\right] J,\cr
J_{{\bf u_3},{\bf R}}&= - \frac 1{18}\left[\left(4 e^{-i \int_{{\bf R}-{\bf v}} A_v} 
                                            + e^{-i \int_{{\bf R}-{\bf w}} A_w} + \left(-2 + 3 e^{-i \int_{{\bf R}} A_{u_3}}\right) \right)\right.\cr 
                     &\ \ \ \ \  \left.  -\sqrt 3 i \left( 4 e^{-i \int_{{\bf R}-{\bf v}} A_v} - e^{-i \int_{{\bf R}-{\bf w}} A_w} + 3 e^{-i \int_{{\bf R}} A_{u_3}} \right)\right] J
\end{align}
defining the Dirac lattice Hamiltonian on the honeycomb lattice for a generic magnetic background in flat space. It is easy to check that by taking ${\bf A} = {\bf K}_D = \frac{4\pi}{3\sqrt 3}(0,1)$ the tunnelings 
$J_{{\bf u_j},{\bf R}}$ become all equal and real as ${\bf K}_D \cdot {\bf v} = - {\bf K}_D \cdot {\bf w} = {\bf K}_D \cdot {\bf u_2}= - {\bf K}_D \cdot {\bf u_3}= \frac{2\pi}3$ and ${\bf K}_D \cdot {\bf u_1} = 0$.
This can be regarded also as an non-trivial check of the validity of \eqref{gaugetun}.

By choosing this pure gauge configuration, it follows that gravitational deformation of the graphene like Hamiltonian in a metric (\ref{metric}) is determined by the hopping
\begin{align}
J_{{\bf u_1},{\bf R}}&= \frac 19\left[\left(J_{{\bf R}- {\bf v}}  
                                            + J_{{\bf R}- {\bf w}}  + 4 J_{{\bf R}}  \right)  -\sqrt 3 i \left(J_{{\bf R}- {\bf v}} - J_{{\bf R}- {\bf w}}  \right)\right],\cr
J_{{\bf u_2},{\bf R}}&= \frac 19\left[J_{{\bf R}- {\bf v}} + 4 J_{{\bf R}- {\bf w}} + J_{{\bf R}} \right],\cr
J_{{\bf u_3},{\bf R}}&= \frac 19\left[4 J_{{\bf R}- {\bf v}} + J_{{\bf R}- {\bf w}} + J_{{\bf R}} \right].
                     \label{Jujdefgra}
\end{align}

\section{Gravitational deformation of the graphene-like Hamiltonian for real and isotropic hopping}\label{sect:IV} 

It is worth to notice the Minkowski metric is not the only one of the form (\ref{metric}) 
associated to an honeycomb tightbinding model with real and equal $J_{\bf u_1}=J_{\bf u_2}=J_{\bf u_3}=f(R)$ at each lattice site. 
In order to systematically analyze the problem, it is convenient to consider a more symmetric formulation  for the hopping (\ref{Jujdefgra}). 
As we can perform the completely equivalent derivation of the discrete Dirac Hamiltonian for ${\bf v},{\bf w}\to -{\bf v},-{\bf w}$, 
the left-right symmetry can be restored by mediating over the two expressions. Explicitly , we find
%
\begin{align}
J_{{\bf u_1},{\bf R}}&= \frac 19\left[\left(\langle J_{{\bf R}}\rangle_{\bf v}  
                                            + \langle J_{{\bf R}}\rangle_{\bf w}  + 4 J_{{\bf R}}  \right)  -\sqrt 3 i \left(\langle J_{{\bf R}}\rangle_{\bf v} - \langle J_{{\bf R}}\rangle_{\bf w}  \right)\right],\cr
J_{{\bf u_2},{\bf R}}&= \frac 19\left[\langle J_{{\bf R}}\rangle_{\bf v} + 4 \langle J_{{\bf R}}\rangle_{\bf w} + J_{{\bf R}} \right],\cr
J_{{\bf u_3},{\bf R}}&= \frac 19\left[4 \langle J_{{\bf R}}\rangle_{\bf v} + \langle J_{{\bf R}}\rangle_{\bf w} + J_{{\bf R}} \right].
                     \label{Jujdefgrasym}
\end{align}
where $\langle J_{{\bf R}}\rangle_{\bf v}\equiv 1/2( J_{{\bf R} + {\bf v}} + J_{{\bf R} - {\bf v}})$, 
and $\langle J_{{\bf R}}\rangle_{\bf w}\equiv 1/2( J_{{\bf R} + {\bf w}} + J_{{\bf R} - {\bf w}})$. 
It is immediate to realize that condition for the tunnelings to be real is
\begin{equation}
\langle J_{{\bf R}}\rangle_{\bf v} = \langle J_{{\bf R}}\rangle_{\bf w}.\label{condJ}
\end{equation}
If we further ask that 
\begin{equation}
\langle J_{{\bf R}}\rangle_{\bf v} = \langle J_{{\bf R}}\rangle_{\bf w}= J_{\bf R},\label{condJ1}
\end{equation}
then the tunnelings in the three directions become equal and proportional to the ``local'' Fermi velocity $J_{\bf R}$ 
\begin{equation}
 J_{\bf u_1}=J_{\bf u_2}=J_{\bf u_3}= \frac 23 J_{{\bf R}}.
  \end{equation}   
  Let me analyze the content of the conditions (\ref{condJ}) and \eqref{condJ1}. 
  Once restricted to smooth  and slowly varying deformations of the Minkowski metric, the former, 
  $\langle J_{{\bf R}}\rangle_{\bf v} = \langle J_{{\bf R}}\rangle_{\bf w}$ forces $\partial_y J(r)=0$, 
  as periodic functions of about one lattice site's period are ruled out by the previous assumption. 
  The condition  \eqref{condJ1} implies linearity, hence $J({\bf r})$ has to be of the Rindler form with $J({\bf r}) \propto (x-x_0)$.
 
Thus, we conclude that whenever the metric can be written in the form \eqref{metric} with a $J({\bf r})=J(x)$, the corresponding tightbinding Hamiltonian on the honeycomb lattice can be made real. 
Generically, the tunnelings will not be isotropic. 
The Dirac Hamiltonian in the Rindler spacetime (see Sect. \ref{sec:Rindler}) provides the only non-trivial instance of a honeycomb tightbinding model with real and isotropic tunnelings, but non constant Fermi velocity.       

\section{Curved spacetimes on  brickwall lattices}\label{sect:V}

I study now a deformed honeycomb lattice. In particular, I stick to the brickwall lattice that is especially simple and suitable for experiments \cite{Tarruell12}. 
The topological structure of the lattice is still the same:
bipartite with coordination number equal to three. Formally, the Hamiltonian has the same expression as in the honeycomb
\begin{equation}
H= \sum_{\bf R} \sum_{j=1}^3 J_{{\bf u_j},{\bf R}}\, b_{{\bf R} + {\bf u_j}}^\dagger a_{\bf R} + {\rm H.c.}, 
\end{equation} 
 but this time the links are  
 \begin{equation}
 {\bf u_1}=(1+\Delta,0), \ \ \ {\bf u_2}=(\Delta,1), \ \ \ {\bf u_3}=(\Delta,-1),\ \ \ |\Delta|\le 1,
 \end{equation}
 which implies that the lattices for $a_{\bf R}$ and $b_{\bf R}$ are square with generators at $\pm\frac{\pi}4$ and the Brillouin zone has the same shape
 with $-(\pi-|k_x|)\le k_y < \pi-|k_x|$ and $-\pi\le k_x < \pi$. 
 

\begin{figure}[hb!]
\resizebox{0.98\columnwidth}{!}{%
\includegraphics{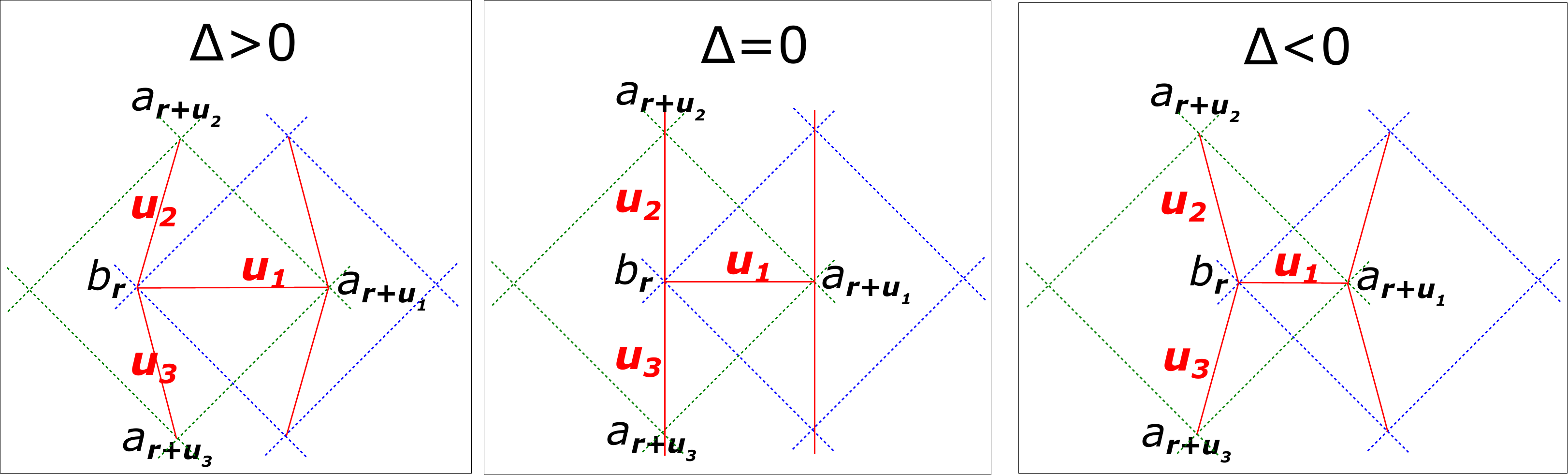}}
\caption{The brickwall lattice, indicated in red, as bipartite lattice formed by two square sublattices, indicated in green and blue, respectively.
The two sublattices are at $45^\circ$, have lattice spacing equal to $\sqrt 2$, and are displaced horizontally one another. 
There are main cases: the displacement (to the right) is less ($\Delta>0$) than, equal ($\Delta=0$) to, or greater ($\Delta<0$) than 1, the length of
the half diagonal of the square lattice. 
A peculiarity of the brickwall lattice is that not all the neighboring sites of the sublattices are connected. In this plot, the missing links are the horizontal 
ones to the right from the sublattice $a$ to the sublattice $b$.  
The allowed links, ${\bf u_j}$, $j=1,2,3$, are given in the main text in terms of $\Delta$.
}
\label{fig:2}       
\end{figure}

 \subsection{Minkowski space}
 
 For constant hoppings $J_{{\bf u_j},{\bf R}}=J_{\bf u_j}$, the existence of Dirac points is easily shown by going in momentum space.
 Defining $a(b)_{\bf R} = \frac 1{2\pi}\int_{BZ} e^{- i {\bf k R}} a(b)_{\bf k}$ we have
 \begin{align}
 H &=\int_{BZ} \sum_j J_{\bf u_j} e^{i {\bf k u_j}} b_{\bf k}^\dagger a_{\bf k} +{\rm H.c.} \cr
   &=  \int_{BZ} \psi_{\bf  k}^\dagger \left\{\left[J_{\bf u_1} \cos ((1+ \Delta) k_x) + (J_{\bf u_2}+J_{\bf u_3})\cos (\Delta k_x)\cos(k_y)\right.\right.\cr
   &   \hspace{2cm}      \left.\left.   - (J_{\bf u_2}- J_{\bf u_3}) \sin (\Delta k_x) \sin(k_y)\right]\sigma_x  \right.\cr 
   &   \hspace{1.5cm}      \left.+  \left[J_{\bf u_1} \sin ((1+\Delta) k_x) 
            + (J_{\bf u_2}+J_{\bf u_3})\sin (\Delta k_x)\cos(k_y) \right.\right.\cr
   &   \hspace{2cm}      \left.\left. + (J_{\bf u_2}- J_{\bf u_3}) \cos (\Delta k_x) \sin(k_y)\right]\sigma_y \right\}\psi_{\bf  k}, \label{eq:gen_mink_brick}
 \end{align}
where $\psi_{\bf  k}\equiv \left(\begin{smallmatrix} a_{\bf k}\\b_{\bf k} \end{smallmatrix}\right)$. 
The location of the Dirac in the BZ is determined by the solution of the system 
\begin{eqnarray*}
&&J_{\bf u_1} \cos ((1+ \Delta) k_x) + (J_{\bf u_2}+J_{\bf u_3})\cos (\Delta k_x)\cos(k_y) - (J_{\bf u_2}- J_{\bf u_3}) \sin (\Delta k_x) \sin(k_y)=0\\ 
&&J_{\bf u_1} \sin ((1+\Delta) k_x) + (J_{\bf u_2}+J_{\bf u_3})\sin (\Delta k_x)\cos(k_y) + (J_{\bf u_2}- J_{\bf u_3}) \cos (\Delta k_x) \sin(k_y)=0.
\end{eqnarray*} 

Let me specialize to the symmetric case $J_{\bf u_j}=J$ and $\Delta=0$. If follows that the two independent Dirac points are located at ${\bf K}_{\pm}=(0,\pm\frac {2\pi}3)$. Note that, differently than 
in the graphene case, the Dirac points seat within the Brillouin zone, which is the square of vertices $(\pm \pi,0)$ and $(0, \pm \pi)$, and not on its borders. Different choices of $\Delta$ and $J_{\bf u_j}$'s lead
to different locations of the Dirac points. For the simple case considered here, the effective Hamiltonian is  $H_\pm=\mp \sqrt 3 p_y \sigma_x + p_x \sigma_y$    which is telling us that the cone is anisotropic, i.e.,  the 
Fermi velocity in the $y$-direction is $\sqrt 3$ times the one in the $x$-direction.
A isotropic cone for $\Delta=0$ is given for instance by the choice $J_{\bf u_2}=J_{\bf u_3}=\frac 1{\sqrt 3} J_{\bf u_1}$. In this case the Dirac points are at $(0,\pm \frac {5\pi}6)$.
An alternative option is  $J_{\bf u_2}=J_{\bf u_3}=\frac 1{\sqrt 2} J_{\bf u_1}$: in this case the  Dirac points are at the boundary of the Brillouin zone, e.g., $(\pi,0)$ and $(0,\pi)$.  
Note that the choice $J_{\bf u_2}=J_{\bf u_3}$ implies that the Dirac points lie on the $y$-axis, for {\it any value} of $\Delta$.

In fact we can get a isotropic cone by construction, i.e.,  by discretizing the isotropic Dirac Hamiltonian on a brick lattice. By setting the Fermi velocity to 1 and by discretizing on the $b$-sites, 
${\bf r} = m({\bf u}_1+ {\bf u}_2) + n ({\bf u}_1- {\bf u}_2)= (m+n,m-n)$, one gets
\begin{equation}
H_{M}= \frac 12 \sum_{\bf r}  \left( i(\partial_x - i \partial_y) a^\dagger_{\bf r}  b_{\bf r} \right)+ {\rm H.c.}\,\label{eq:mink_ham},
\end{equation}                                                          
where the discrete derivatives are
\begin{align}
\partial_x a_{\bf r} &= a_{{\bf r} +{\bf u}_1} - \frac 12 ( a_{{\bf r} +{\bf u}_2} + a_{{\bf r} -{\bf u}_2}), \cr
\partial_y a_{\bf r} &= \frac 12 ( a_{{\bf r} +{\bf u}_2} - a_{{\bf r} -{\bf u}_2}). \label{discretederiv}
\end{align}     
The substitution gives
\begin{equation}
H_{M}= \frac 12 \sum_{\bf r} \left( \left( e^{i\frac {\pi}2 } a^\dagger_{{\bf r}+{\bf u}_1} + \frac{e^{-i \frac 34 \pi} a^\dagger_{{\bf r}+{\bf u}_2}+e^{-i \frac {\pi}4 } a^\dagger_{{\bf r}+{\bf u}_3}}{\sqrt 2}\right) 
    b_{\bf r} \right) + {\rm H.c.},
\end{equation} 
which, up to the gauge transformation $c_{\bf s} \to e^{i \frac {\pi}4 ({\bf s}\cdot {\bf u}_2 + 2)} c_{\bf s}$ (or equivalently, $c_{(x,y)} \to e^{i \frac {\pi}4 (y + 2)} c_{(x,y)}$), 
where $c_{\bf s}$ are both $a_{\bf s}$ and $b_{\bf s}$, is equivalent to the Hamiltonian \eqref{eq:gen_mink_brick} for $\Delta=0$ and $J_{\bf u_1}=1=\sqrt 2 J_{\bf u_2}=\sqrt 2 J_{\bf u_3}$.

\subsection{Rindler space} \label{sec:Rindler}

\begin{figure}\sidecaption
\includegraphics[width=9cm]{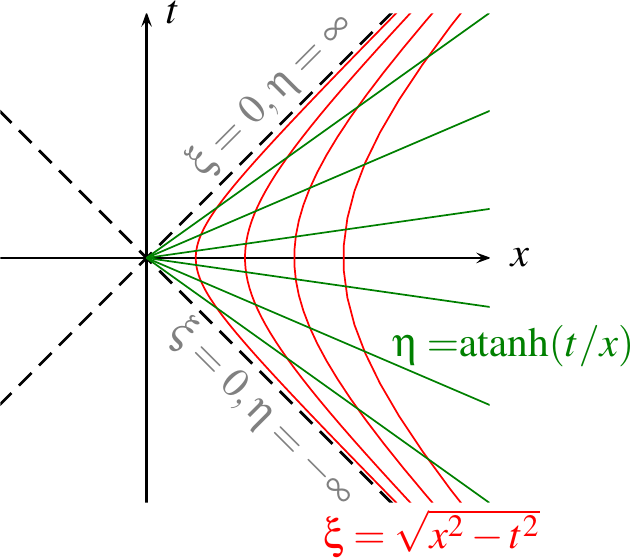}
\caption{Rindler coordinates on 1+1D Minkowski spacetime, $\eta$
  (Rindler time) and $\xi$ (Rindler space). The Rindler wedge,
  delimited by dashed lines, is the domain of validity of the
  coordinate patch. The dashed lines represent 
  the horizon, i.e., the loci of infinity acceleration.
  Constant $\eta$ lines (green) are spacelike, and
  constant $\xi$ lines (red) are timelike. For simplicity, we plot the
  trajectories only in the right wedge, $x>0$, as the ones for the
  wedge $x<0$ can be obtained by reflection around the $y$-axis.   
  }
\label{fig:rindler}
\end{figure}

Now I repeat the above construction for the Dirac Hamiltonian in 2+1 Rindler spacetime. 
Rindler spacetime is Minkowski spacetime
viewed by an accelerated observer \cite{Misner,Wald,Sachs}. 
In special relativity, an observer moving with constant acceleration follows an 
hyperbolic trajectory. For a unitary acceleration in natural unit,  
 (for
convenience in the following we take the speed of light to be $c=1$)
in the positive $x$-axis, the trajectory  
in the parametric form reads  
\begin{equation}
\begin{cases}
t=\xi \sinh \eta\\
x=\xi \cosh \eta
\end{cases}.
\label{rindler.coord}
\end{equation}
The parameter $\eta$ plays the role of the co-moving time coordinate for this observer, and $\xi$ of
the co-moving space coordinate. They are called {\em Rindler
  coordinates}, and are related to the Minkowski Cartesian ones  by \eqref{rindler.coord} (see
Fig. \ref{fig:rindler}). The accelerated observer is at rest in Rindler spacetime.
 Notice the similarity with polar coordinates, where $\xi$
plays the role of a radius and $\eta$ is an angle in hyperbolic
geometry. The principle of equivalence states that physics seen by a
non-inertial observer can be absorbed by a change in her
metric. Indeed, in these coordinates, the Minkowski metric becomes
\begin{equation}
ds^2 = - \xi^2 d\eta^2 + d\xi^2 + dy^2,
\label{rindler.metric}
\end{equation}
which is known as the Rindler metric. It is of the form of \eqref{metric} (once we rename
$\xi$ with $t$ and $\xi$ with $x$), with a function $J$ linear in $\xi$. 
Notice that the Rindler time
direction corresponds to a symmetry of the metric, i.e., it
constitutes a Killing vector which is inequivalent to the usual
Minkowski time direction. In fact, it corresponds to a boost
transformation. In the polar coordinates view, it is the generator of
hyperbolic rotations. Another peculiarity of a relativistic constant acceleration,
which is reflected by the hyperbolic geometry, is that the back of a rigid stick oriented along $x$ 
has to accelerate more than its front. In fact, a 
static observer in Rindler spacetime feels a proper acceleration that it is inversely proportional to 
$\xi$. The acceleration diverges at $\xi=0$ that corresponds to a
singularity in the coordinate system, singularity that is signaled in the Rindler metric \eqref{rindler.metric} 
by the vanishing of $d\eta^2$ coefficient. This is the hallmark of an {\em event
  horizon}. Thus, spacetime is separated into two parts which do not
communicate: the two Rindler wedges, $\xi>0$ and $\xi<0$. 
The existence of an event horizon is at the heart of many intriguing phenomena
like the Unruh effect and the appearance of the Hawking radiation from black holes. In fact, 
the Rindler metric describes the near-horizon limit of a Schwartzschild  black hole.
  
By renaming the Rindler coordinates by $t$ and $x$, the continuous Rindler Hamiltonian in Rindler spacetime reads   
\begin{equation}
H_{R}= \frac 12 \sum_{\bf r} |{\bf r}\cdot {\bf u}_1| \left( i(\partial_x - i \partial_y) a^\dagger_{\bf r}  b_{\bf r} \right) + {\rm H.c.},
\end{equation}                                                          
where the discrete derivatives are taken as in \eqref{discretederiv}. 

After the substitution we have
\begin{multline}
H_{R}= \frac 12 \sum_{m,n}  
\left( \left( |m+n +\frac 12|\,e^{i\frac {\pi}2 }a^\dagger_{m+n+1,m-n}\right.\right.\cr
\left.\left. +\,|m+n| \,  \frac{e^{-i \frac 34 \pi} a^\dagger_{m+n,m-n+1}+e^{-i \frac {\pi}4 } a^\dagger_{m+n,m-n-1}}{\sqrt 2}\right)  b_{m+n,m-n} \right)+ {\rm H.c.},
\end{multline} 
where the value of the warp factor $|{\bf r}\cdot {\bf u}_1|$ is averaged over the corresponding link, such to give rise to anisotropy in the link intensity.
The phases can be reabsorbed as in the Minkowski case and the Hamiltonian cast in a real form 
\begin{multline}
H_{R}= \frac 12 \sum_{m,n}  \left( \left( |m+n +\frac 12|\, a^\dagger_{m+n+1,m-n} \right.\right.\cr
\left.\left. +\,  |m+n| \,  \frac{a^\dagger_{m+n,m-n+1}+ a^\dagger_{m+n,m-n-1}}{\sqrt 2}\right)  b_{m+n,m-n} \right) +  {\rm H.c.}\,. \label{eq:rindler_brick_H1}
\end{multline} 
This is a possible choice for the implementation, up to an overall energy scale that fixes the ``local'' speed of light.

An alternative implementation can be obtained by ``rotating the brick of $90^{\circ}$'' (in fact it is a reflection) and defining ${\bf u}_1= (0,1)$, and ${\bf u}_2=-{\bf u}_3= (1,0)$. 
Accordingly, ${\bf r}= (m-n,m+n)$. Repeating the same exercise as before we have
\begin{equation}
H_{R}= \frac 12 \sum_{\bf r} |{\bf r}\cdot {\bf u}_2| \left( i(\partial_x - i \partial_y) a^\dagger_{\bf r}  b_{\bf r} \right)+ {\rm H.c.}\,,
\end{equation}                                                          
where the discrete derivatives with respect to $x$ and $y$ are just exchanged
\begin{align}
\partial_x a_{\bf r} &= \frac 12 ( a_{{\bf r} +{\bf u}_2} - a_{{\bf r} -{\bf u}_2}), \cr
\partial_y a_{\bf r} &= a_{{\bf r} +{\bf u}_1} - \frac 12 ( a_{{\bf r} +{\bf u}_2} + a_{{\bf r} -{\bf u}_2}). 
\end{align} 
After the gauge transformation  $c_{\bf s} \to e^{i \frac {\pi}4 {\bf s}\cdot {\bf u}_2} c_{\bf s}$, the final Hamiltonian reads
\begin{multline}
H_{R}= \frac 12 \sum_{m,n}  \left( \left( \frac{|m-n +\frac 12|\, a^\dagger_{m-n+1,m+n}+ |m-n -\frac 12|\, a^\dagger_{m-n-1,m+n}}{\sqrt 2}   \right.\right.\cr
\left.\left. +\,  |m+n| \, a^\dagger_{m-n,m+n+1}\right)  b_{m-n,m+n} \right)+ {\rm H.c.}
\,. \label{eq:rindler_brick_H2}
\end{multline}  

\subsection{Dirac Hamiltonian in Rindler space with extradimensions} 

The above Hamiltonians \eqref{eq:rindler_brick_H1} and \eqref{eq:rindler_brick_H2} can be conveniently implemented in a synthetic lattice. 
For synthetic lattice, I mean a lattice in which at least one of the spatial dimensions is obtained by promoting some internal degrees of freedom of the constituents
 --here the spin states for atoms-- to sites of an extradimension \cite{Boada12}, also known as synthetic dimension \cite{Celi14}. The tunneling in the synthetic dimension is 
 induced by some coherent coupling of the internal states. For atoms, such couplings can be induced by radiofrequency or by Raman pulses. The key advantage of synthetic lattices
 is that they are very versatile \cite{Rodriguez16} and convenient experimentally, as demonstrated in \cite{Stuhl15,Mancini15}, see also \cite{Celi15}.        

As I am interested to have a synthetic lattice extended in the $x$-direction 
(perpendicular to the horizon), I implement the $y$-direction in the synthetic dimension by using atomic spin states. I indicate the spin states with an index $\sigma$ and the position along the 
chain with an index  $l$. Note that on the sites occupied by the $b$-fermions, even (odd) values of $l$ correspond to even (odd) values of the spin $\sigma$, while for the $a$-sites exactly the opposite occurs, 
even (odd) $l$ correspond to odd (even) $\sigma$. Thus, by summing over all $l=-L,\dots,L$, I can use just one species with $S$ internal states. 
   In case of \eqref{eq:rindler_brick_H1}, we have 
 \begin{equation}
H_{R}= \frac 12 \sum_{l,\sigma}  
\left( \left(\sqrt 2 \, |l|\,(1-\frac{\delta_{\sigma,1}+\delta_{\sigma,S - 1}}2) c^{\dagger(\sigma+1)}_{l}\,   + \, | l + \frac{\epsilon(l,\sigma)}2|\,  c^{\dagger(\sigma)}_{l+\epsilon(l,\sigma)}\right)c^{\sigma}_{l} \right) 
+  {\rm H.c.}\,, \label{eq:rindler_brick_Hextradim1}
\end{equation}   
where we have defined $\epsilon(l,\sigma)\equiv(-1)^{l+\sigma}$. Note that due to the definition of the discrete derivative in the $y$-direction, Eq. \eqref{discretederiv}, the tunneling terms along $y$ have 
on the boundary of the
synthetic dimension half of the strength than in the bulk. Such result is directly obtained from the Hamiltonian \eqref{eq:rindler_brick_H1} by collecting in a unique ``forward'' 
tunneling terms in $y$ the forward and back tunneling terms in \eqref{eq:rindler_brick_H1}. 

   In case of \eqref{eq:rindler_brick_H2}, {\it which is probably the easiest to be simulated}, we have 
 \begin{equation}
H_{R}= \frac 12 \sum_{l,\sigma}  \left( \left(\sqrt 2 \, |l+\frac 12|\,(1-\frac{\delta_{l,-L}+\delta_{l,L - 2}}2) c^{\dagger(\sigma)}_{l+1}\,   
                                               + \, | l|\,  c^{\dagger(\sigma+\epsilon(l,\sigma))}_l \right)c^{\sigma}_{l}
                                   \right) +  {\rm H.c.}\,. \label{eq:rindler_brick_Hextradim2}
\end{equation}
A similar reasoning as above (with the interchange of $y$ with $x$) explains the boundary terms, this time in the real dimension.  


It is worth to notice that the Hamiltonian \eqref{eq:rindler_brick_Hextradim2} once specialized to a one-dimensional chain describes correctly the tightbinding Dirac Hamiltonian in 1D. As the 
$$H_{R-1D} = \frac 12 \sum_{\bf r} |{\bf r}\cdot {\bf u}_1| \left( i\partial_x  a^\dagger_{\bf r}  b_{\bf r} \right) + {\rm H.c.},$$   
the final expression,  a part a factor $\sqrt 2$ will correspond to \eqref{eq:rindler_brick_Hextradim2} for S=1, i.e.
\begin{equation}
H_{R-1D}= \frac 12 \sum_{l}  \left(   |l+\frac 12|\,(1-\frac{\delta_{l,-L}+\delta_{l,L - 2}}2) c^{\dagger}_{l+1}\,   
                                               c_{l}
                                   \right) +  {\rm H.c.}\,. \label{eq:rindler_brick_Hextradim21D}
\end{equation}
 
\section{$\pi$-flux Hamiltonian}\label{sect:VI}

The most elegant way of deriving the Dirac Hamiltonian in the $\pi$-flux form is to begin with a bipartite square lattice.
As usual I start from \eqref{eq:mink_ham}. I take ${\bf u}_1= (1,0)$ and ${\bf u}_2= (0,1)$.
Then, the fact that I want links turned on in any direction is telling me that all the neighbors 
$a_{{\bf r}+{\bf u}_1},a_{{\bf r}+{\bf u}_2},a_{{\bf r}-{\bf u}_1},a_{{\bf r}-{\bf u}_2}$ of $b_{\bf r}$ have to enter in the discrete derivative
of $a_{\bf r}$. A simple choice is
\begin{align}
\partial_x a_{\bf r} &= \frac{a_{{\bf r} +{\bf u}_1} - a_{{\bf r} -{\bf u}_1}}2, \cr
\partial_y a_{\bf r} &= \frac{a_{{\bf r} +{\bf u}_2} - a_{{\bf r} -{\bf u}_2}}2. 
\end{align} 
After substituting the discrete derivatives we get
\begin{equation}
H_{M}= \frac 14 \sum_{\bf r} \left(\left( e^{i\frac {\pi}2 } a^\dagger_{{\bf r}+{\bf u}_1} + e^{-i\frac {\pi}2 } a^\dagger_{{\bf r}-{\bf u}_1}
                                   + a^\dagger_{{\bf r}+{\bf u}_2}-  a^\dagger_{{\bf r}-{\bf u}_2}\right) 
    b_{\bf r} \right) + {\rm H.c.},
\end{equation}
    which, by applying the gauge transformation $c_{\bf s} \to e^{i\frac {\pi}2 {\bf s}\cdot {\bf u}_1 } c_{\bf s}$, is equivalent to
\begin{equation}
H_{M}= \frac 14 \sum_{\bf r} \left(\left( a^\dagger_{{\bf r}+{\bf u}_1} + a^\dagger_{{\bf r}-{\bf u}_1}
                                   + a^\dagger_{{\bf r}+{\bf u}_2}-  a^\dagger_{{\bf r}-{\bf u}_2}\right) 
    b_{\bf r} \right) + {\rm H.c.}\,.
\end{equation}
By using just a single species the above Hamiltonian can be cast in the more familiar form
\begin{equation}
H_{\text{M $\pi$-flux}}= \frac 12 \sum_{m,n} \left(\left( c^\dagger_{m+1,n} + (-1)^{m+n} c^\dagger_{m,n+1}\right) 
    c_{m,n} \right) + {\rm H.c.}\,.
\end{equation}
   
The step to Rindler Hamiltonian is extremely easy: we have to include the dependence of hopping strength on the distance from the horizon, for instance, measured from
the center of the link. By placing the horizon at $m=0$ we have
\begin{equation}
H_{\text{R $\pi$-flux}}= \frac 12 \sum_{m,n} \left(\left( |m+\frac 12|\, \,  c^\dagger_{m+1,n} + (-1)^{m+n} |m|\, c^\dagger_{m,n+1}\right) 
    c_{m,n} \right) + {\rm H.c.}\,.
\end{equation}
It is worth to note that it is immediate to promote the position index $n$ to a spin label and to implement the direction parallel to the horizon as
a synthetic dimension. Let me just comment that in order to simulate the Unruh effect we proposed in \cite{Rodriguez16} to implement a similar Hamiltonian to the one above,
but with the $\pi$ flux realized in the symmetric gauge, as  it is easier to achieve in optically shaken in 2D real lattices (cf. \cite{Aidelsburger.13,Miyake.13,Kennedy.15}).  


\section{Conclusions \& Outlook}\label{sect:VII}

In this paper, I have illustrated the power of our top-down approach:   
 formulating the Dirac Hamiltonian in curved spacetime directly in position space allows for the construction of several models 
 of emerging gravitating Dirac fermions that can be simulated in optical lattices.
 It is worth to notice that this approach is in principle useful in any dimensions, and it could be extended to time-dependent metrics. 
 Furthermore, as top-down approach allows the choice of the lattice and, thus, of its properties under parity symmetry,
it can be used for obtained lattice formulation of topological insulators in specific classes in generic dimension as we 
do in \cite{Bikash} for 3D models. 
 
Here I have considered naive Dirac Hamiltonian as doubling does not play a major role, for instance, in the Unruh effect.
Obviously, doubling could be avoided,  for instance,  by including generalization 
of Wilson or domain wall fermions \cite{Rubakov83,Callan85} to curved spacetime.
Such research direction may be interesting both for testing which properties of known topological models are altered by coupling to gravity
(cf. \cite{Can14}), or for considering interacting gravitating fermions. 
The former direction requires the introduction of ``mass'' terms  that are compatible with the Dirac Hamiltonian in curved spacetimes and can open
a gap, as it happens in Minkowski spacetime. Such terms are not discussed here but they can be simply achieved by making the ordinary mass terms in Minkowski spacetimes, e.g, 
the staggered chemical potential for the $\pi$-flux formulation of the tightbinding Dirac Hamiltonian, position dependent. For the optical metrics of Eq. \eqref{metric}, such dependence 
is determined by the local Fermi velocity $J({\bf r})$, as it happens for the tunneling terms.    

The possibility of considering strongly interacting gravitating matter, and perhaps the possibility
of including matter backreaction on the artificial metric via density-dependent hopping \cite{Dutta15} are very appealing. They are unique and distinctive features of our quantum simulation strategy based on
ultracold fermionic atoms in optical lattices developed in \cite{Boada2010b} and \cite{Rodriguez16}, and applied here, features that distinguish it from other analogue gravity approaches.

\begin{acknowledgement}
This work has been supported by Spanish MINECO (SEVERO OCHOA Grant
SEV-2015-0522, FOQUS FIS2013-46768, and FISICATEAMO FIS2016-79508-P), 
the Generalitat de Catalunya (SGR 874 and CERCA program),
Fundaci\'o Privada Cellex, and EU grants EQuaM (FP7/2007-2013 Grant
No. 323714), OSYRIS (ERC-2013-AdG Grant No. 339106), SIQS
(FP7-ICT-2011-9 No.  600645), QUIC (H2020-FETPROACT-2014 No. 641122)
and PCIG13-GA-2013-631633. 
\end{acknowledgement}


\end{document}